\documentclass[10pt]{iopart}
\usepackage{graphicx,rotating,subfigure} 
\usepackage{iopams}  
\usepackage{dsfont}

\usepackage{graphicx}
\usepackage{dcolumn}
\usepackage{bm}
\usepackage{epsfig}

\newcommand{\be}{\begin{eqnarray}}
\newcommand{\ee}{\end{eqnarray}}
\newcommand{\xop}{\mathbf{b}}
\newcommand{\yop}{{\mathbf{b}^\dagger}}
\newcommand{\Nop}{\mathbf{n}}
\def\refeq#1{(\ref{#1})}
\def\d{\mbox d}
\def\wt{\widetilde}
\def\nn{\nonumber}

\def\ip{\int_{0}^{\infty}}

\def\S{\mathcal S}

\def\la{\lambda}

\def\e{\epsilon}

\def\ve{\varepsilon}
\def\l{\left}
\def\r{\right}
\def\te{\mbox{e}}
\def\rmi{{\rm i}}
\def\re{\mbox{Re }}

\def\tr{\mbox{tr}}

\begin{document}
\bibliographystyle{jpa}
\title{Finite temperature Drude weight of an integrable Bose chain}
\author{Michael Bortz}
\address{Department of Theoretical Physics, Research School of Physics and Engineering, Australian National University, Canberra ACT 0200, Australia}
\begin{abstract}
We study the Drude weight $D(T)$ at finite temperatures $T$ of an integrable bosonic model where the particles interact via nearest-neighbour coupling on a chain. At low temperatures, $D(T)$ is shown to be universal in the sense that this region is equivalently described by a Gaussian model. This low-temperature limit is also relevant for the integrable one-dimensional Bose gas. We then use the thermodynamic Bethe ansatz to confirm the low-temperature result, to obtain the high temperature limit of $D(T)$ and to calculate $D(T)$ numerically. 
\end{abstract}
\section{Introduction}
The one-dimensional Bose gas with $\delta$-interaction was solved exactly by Lieb and Liniger \cite{ll63}; thermodynamic properties were calculated by Yang and Yang \cite{yy69}, using a method which is nowadays known as thermodynamic Bethe ansatz (TBA). This Bose gas model can be viewed as the continuum limit of a one-dimensional lattice model, where the particles interact via a short-range interaction. The advantage of a lattice description is that integrals over distances (momenta) have a natural lower (higher) cutoff, namely the lattice constant (bandwidth). This makes it possible to calculate high-energy properties, in our case, the high-temperature limit of $D(T)$. In the other extreme, at low energies, a continuum description of the lattice model is justified, which opens the possibility of covering the low-energy regime of the continuum model as well.

Here, we want to benefit from the existence of an integrable lattice regularization of the Lieb-Liniger (LL) Bose gas, stemming from a $q$-deformation of the underlying bosonic commutators. This $q$-Bose model was shown to be integrable in \cite{bog92,bog93,bul95}, where the close relation to the Lieb-Liniger model in the continuum limit was also pointed out. Very recently \cite{borser06}, the lattice model has been rederived in the framework of the algebraic Bethe ansatz. In that approach, the particle current as a conserved quantity appeared naturally. 

In this work, we study the dynamical conductivity $\sigma(\omega,T)$ of the $q$-Bose model, depending on the frequency $\omega$ and the temperature $T$. It is defined by 
\be
\mathcal J(\omega,T)=\sigma(\omega,T) E (\omega)\label{conddef},
\ee
where $\mathcal J$ is the particle current induced by the field $E$. The particle current is conserved \cite{borser06}, leading to 
\be
\re \sigma(\omega,T)=\pi  \delta(\omega) D(T)\label{resig},
\ee
that is, the conductivity is infinite at zero frequency. Our aim here is to calculate the temperature dependence of the Drude weight $D(T)$. 

This paper is organized as follows: In section 2, we shortly review the model and provide some background on the calculation of $\sigma(\omega,T)$. The third section is devoted to the calculation of the leading term in $D(T=0)$ from the effective low-energy Gaussian field theory. Section 4 contains the calculation of $D(T)$ from the TBA. The $T=0$-result of the preceding section is recovered here, as well as an analytical formula in the high-temperature region $D(T\gg 1)$. Numerical results for a wider range of temperatures interpolate between these two extremes. We end with a conclusion and outlook in section 5. 

\section{Conserved current in the $q$-Bose model}
In this section, we shortly review the $q$-Bose model following \cite{borser06}. Let $\xop,\yop,\Nop$ be the generators of the $q$-oscillator algebra:
\be
\xop\yop=1-q^{2\Nop+2}\;,\quad \yop\xop = 1-q^{2\Nop}\;,\nn\\
\quad \xop
\Nop=(\Nop+1)\xop\;,\quad \yop \Nop=(\Nop-1)\yop\;,\nn
\ee
with $0\leq q<1$. In the following, it is often convenient to use the parameterization
\be
q=\te^{-\eta},\; \eta\in\l]0,\infty\r[\nn.
\ee
A $q$-oscillator is represented in the Fock space:
\begin{equation}\label{Fock}
\Nop|n\rangle=n|n\rangle\;; \quad n = 0,1,2,3,\dots\;;\;
\xop|0\rangle =0\;.
\end{equation}
We deal with a chain of length $L$ where the local $q$-oscillator
algebra is assigned to each site $\ell=1,\ldots,L$. The quantities 
\be
\mathcal{N}&=&\sum_{\ell=1}^L \Nop_\ell\nn\\
\mathcal{P}_+&=&\frac{1}{1-q^2}\sum_{\ell=1}^L \xop_{\;\ell\;}^{} \yop_{\;\ell+1\;}^{}\;,\mathcal{P}_-=\frac{1}{1-q^2} \sum_{\ell=1}^L
\xop_{\;\ell\;}^{} \yop_{\;\ell-1\;}^{}\;,\nn
\ee
commute pairwisely. The eigenvalue of $\mathcal N$ is the number of particles $N$. The choice of the Hamiltonian
\be
\mathcal{H}\;=\;- \frac{1}{2} (\mathcal{P}_+ + \mathcal{P}_-)-\mu
\mathcal{N}\,, \label{defh}
\ee
where $\mu$ is the chemical potential, leads \cite{bog92,bog93,bul95} in the continuum limit to the Lieb-Liniger Bose gas with Hamiltonian
\be
\fl \mathcal H= \frac{\Delta}{2} \int \l[ \partial_x \Psi^\dagger(x) \partial_x \Psi(x)+  2c \Psi^\dagger(x)\Psi^\dagger(x)\Psi(x)\Psi(x)-\mu\Psi^\dagger(x)\Psi(x)\r]\d x \label{nls}.
\ee
Here $\Delta$ is the lattice constant of the $q$-Bose model and the coupling constant $c$ has been introduced such that $\eta=c\cdot \Delta$. Furthermore, trivial constants have been absorbed into $\mu$. 

 The current operator $\mathcal J$ is defined by 
\be
\mathcal J&=& \sum_{\ell=1}^L j_\ell\nn\\
\partial_t \Nop_\ell&=&j_{\ell-1} -j_\ell\nn,
\ee
where $\partial_t \Nop_\ell\equiv \rmi \l[\mathcal{H},\Nop_\ell\r]$. It turns out that 
\be
j_\ell=\xop_\ell \yop_{\ell +1} - \xop_{\ell +1} \yop_\ell \label{jl},
\ee
so that $\mathcal J= \frac{1}{2\rmi} \l( \mathcal{P}_+ - \mathcal{P}_-\r)$,
and thus 
\be
\l[ \mathcal H,\mathcal J\r] = 0\label{com},
\ee
i.e. $\mathcal J$ is a conserved quantity.

Let us finally write the current operator in the same continuum limit that lead from \refeq{defh} to \refeq{nls}:
\be
\mathcal J&=& \frac{1}{2\rmi} \,\int \l\{\Psi(x) \partial_x \Psi^\dagger(x) - \l[\partial_x \Psi(x)\r] \Psi^\dagger(x)\r\}\,\d x\nn,
\ee
which coincides \cite{korbook} with the total momentum operator of the model \refeq{nls}. This equality of the total current and momentum operator is, however, not true for the lattice model \cite{che06}. 

Within linear response theory, the current is proportional to the field, where the constant of proportionality is the conductivity. It is related to the current-current correlation function (Kubo-formula, \cite{kub57,mah00}) as follows:
\be
\sigma(\omega,T)&=& \ip \,\d t\,\te^{\rmi \omega t} \int_{0}^\beta \,\d \tau\,\langle \mathcal J(-t-\rmi \tau) \mathcal J\rangle \nn, 
\ee
where $t$ is the (real) time, $\beta:=1/T$,  and $\langle\cdots\rangle$ denotes the thermal average. Because of Eq.~\refeq{com} we have
\be
\sigma(\omega,T)&=& \lim_{\e\to 0}\frac{\beta}{\rmi(\omega-\rmi \e)} \langle \mathcal J^2\rangle\nn,
\ee
from which Eq.~\refeq{resig} follows with
\be
D(T)=\beta \langle \mathcal J^2\rangle \label{dt}.
\ee
The latter expectation value is obtained straightforwardly from 
\be
\langle \mathcal J^2 \rangle &=& \partial^2_\lambda \ln \tr \l.\te^{-\beta \mathcal H - \lambda \mathcal J}\r|_{\lambda=0}\label{jsqu}.
\ee

As mentioned above, the model defined by the Hamiltonian \refeq{defh} is Bethe ansatz (BA) solvable. Because of Eq.~\refeq{com}, this is also true for 
\be
\wt{\mathcal H}&:=& \mathcal H - \tilde \lambda \mathcal J\label{wth},
\ee
where $\tilde \la=\la/\beta$. Following the sketch of the BA solution in \cite{borser06}, it is not difficult to show that the corresponding eigenvalues $\wt E$ read
\be
\wt E&=& \sum_{j=1}^N \tilde \ve(k_j)\nn\\
\tilde \ve(k)&=& -\cos k - \tilde \lambda \sin k -\mu\nn,
\ee
where the quantum numbers (BA-roots) $k_n$ are to be determined from the set of coupled algebraic equations 
\be
\te^{\rmi k_j L}&=& \prod_{m\neq j} \frac{\sin\l[(k_j-k_m)/2 + \rmi \eta\r]}{\sin\l[(k_j-k_m)/2 - \rmi \eta\r]},\; j=1,\ldots,N\label{bae}.
\ee
In the thermodynamic limit, ground state and thermodynamic properties of $\mathcal H$ have been determined from these equations \cite{bog92,bog93,bul95,borser06}. In section 4, the Drude weight $D(T)$ will be calculated along similar lines. Let us finally note that the continuum limit which lead to the bose gas \refeq{nls} corresponds to the rational limit $k_n= p_n \cdot\Delta$, $\eta=c\cdot \Delta$, $L=\tilde L/\Delta$ in \refeq{bae} at fixed particle number $N$, where $\tilde L$ is the length of the system. This limit leads to the BA equations found by Lieb and Liniger \cite{ll63}.  

\section{The low-energy effective theory}
In this section, an effective field theory is constructed for the model \refeq{wth} in order to calculate the leading term in a low-temperature expansion of $D(T)$.

A density-phase representation of the operators $\xop, \yop$ has been used in
\cite{borser06} to arrive at a Gaussian model which describes the low-energy,
i.e. low-temperature properties of the model \refeq{defh}. In the field-theoretical calculation we also restrict ourselves to a deformation parameter $q$ which is close to 1, i.e. $\eta \gtrsim 0$. However, the final results will be valid for the whole range $\eta\in\l]0,\infty\r[$. We will come back to this point at the end of this section. 

Following \cite{borser06}\footnote{We use a slightly different notation
  here.}, a continuum limit different from the one leading to \refeq{nls} is
performed. Therefore we set $x=\ell\cdot \Delta$, $\Nop_\ell=\rho(x)\cdot
\Delta$, where $\Delta$ is the lattice constant. Using a density-phase
representation of the bosonic operators $\xop_\ell, \yop_\ell$, one obtains an
  infinite series, of which we only give the leading terms here:
\be
\frac{\xop_\ell}{\sqrt{\Delta}}= \xop(x) &=&  \sqrt{2\eta
  \rho(x)}\l(1-\frac{\rho(x) \eta \Delta}{2}+ \mathcal{O}\l(\l(\rho \eta \Delta\r)^2\r)\r)\te^{\rmi \theta(x)}\label{xopapp}\\
\frac{\yop_\ell}{\sqrt{\Delta}}= \yop(x) &=& \te^{-\rmi \theta(x)}\sqrt{2\eta  \rho(x)}\l(1-\frac{\rho(x) \eta \Delta}{2}+ \mathcal{O}\l(\l(\rho \eta \Delta\r)^2\r)\r)\label{yopapp}.
\ee
 Furthermore, we split $\rho(x)=n+\Phi(x)$ where $n$ is the average particle
  density per unit length.\footnote{In the framework of the BA solution, the
  same symbol is used for the average particle density per lattice site.} The
  density operator $\rho$ is canonical conjugate to the phase operator
  $\theta$, such that $\l[\Phi(x),\theta(y)\r]=\rmi \delta(x-y)$. Since $\Phi(x)$ has been introduced to encode the low-energy density fluctuations, we require $\langle \Phi(x)\rangle \ll n,\;\langle \partial_x \Phi(x)\rangle \ll 1$. To gain insight into the physical significance of $\theta(x)$, we observe that, following \refeq{jl}, the local current density $J(x)$ is written as $J(x)=n\partial_x \theta(x)$, so that $\partial_x\theta(x)$ stands for the local momentum fluctuations.    

Under rescaling 
\be
\theta=\theta'/\sqrt{K_c}\,, \;\Phi= \sqrt{K_c} \Phi'\,,\label{scal}
\ee
where the charge velocity $v_c$ and the Luttinger parameter $K_c$ are, in this approximation, given by ($\eta=c\Delta\ll 1$)
\be
v_c=\sqrt{ 2nc}\,,\;K_c=\sqrt{\frac{n}{2c}}\label{vck}\,,
\ee
the Hamiltonian density describing the low-energy fluctuations is written as
(we discard a trivial additive constant)
\be
H_0 &=& \frac{v_c}{2}\l[ \l(\partial_x \theta'\r)^2 + \l(\Phi'\r)^2\r]\label{h0}\\
\wt H&=&H_0-\mu \sqrt{K_c} \Phi'- \tilde \la \frac{n}{\sqrt{K_c}} \partial_x \theta'\nn.
\ee
Let us  write $n=v_c\cdot K_c$ in the last term so that the only parameters that enter in the Hamiltonian are $v_c,K_c$ and the potentials $\mu,\, \tilde \la$:
\be
\wt H=H_0-\mu \sqrt{K_c} \Phi'- \tilde \la v_c \sqrt K_c \partial_x \theta'\label{gauss2}.
\ee
From the Lagrangian $L_0$ corresponding to $H_0$, the equation of motion 
\be
\partial_t^2 \theta'/v_c - v_c \partial_x^2 \theta'=0\label{eom}
\ee
is obtained, where $\partial_t \theta'/v_c=\Phi'$. Eq.~\refeq{eom} can be interpreted as a continuity equation for a current $v_c \partial_x \theta'$ and a density $\partial_t \theta'/v_c=\Phi'$ (the existence of a conserved quantity with these current and density is due to the invariance of the Lagrangian under translations $\theta\to \theta + a$). This is consistent with our interpretations of $\theta$ and $\Phi$ above.   

From \refeq{gauss2}, we can infer the zero-temperature charge susceptibility $\chi_c$ and current-current-correlation function $\langle \mathcal J^2\rangle$. Namely, upon shifting $\Phi'\to \Phi'+\frac{\mu \sqrt{ K_c}}{v_c}$, $\partial_x\theta'\to \partial_x \theta'+\tilde\la \sqrt{ K_c}$ we obtain
\be
\wt H=H_0-\frac{\mu^2 K_c}{2 v_c}- \frac{\tilde \la^2  v_c K_c}{2} \nn.
\ee
Thus
\be
\chi_c(T=0)= \frac{K_c}{v_c}\,,\qquad \lim_{T\to 0}\langle \mathcal J^2\rangle/T= v_c K_c\label{chic},
\ee
so that
\be
D(T= 0)=v_c K_c \label{dsmallt}\,.
\ee

The Gaussian model \refeq{gauss2} is written in the standard form in terms of left- and right-moving fields $\phi_{L,R}$ by introducing $\phi'$ such that $\Phi'=\partial_x \phi'$ and $\phi'=\phi_L+\phi_R$, $\theta'=\phi_L-\phi_R$, \cite{caz04}. Then
\be
\fl\wt H = v_c \l[ \l( \partial_x \phi_L\r)^2 + \l( \partial_x \phi_R\r)^2 \r] -\mu \sqrt{ K_c} \partial_x\l(\phi_L + \phi_R\r) - \tilde\la v_c \sqrt{ K_c} \partial_x \l( \phi_L-\phi_R\r)\label{lrmod}.
\ee
A few comments are in order at this point. First of all, Eqs.~\refeq{chic}, \refeq{dsmallt} have been derived for $\eta\to 0$, resulting in \refeq{vck}. However, as will be shown in the next section, Eq.~\refeq{dsmallt} is valid for the whole range $\eta\in \l]0,\infty\r[$, where $v_c,\,K_c$ have to be calculated from the exact solution (the same is true for $\chi_c$ given in Eq.~\refeq{chic}, as shown in \cite{borser06}). In the two special cases $\eta\to0$, $\eta\to\infty$, explicit results are available:
\be
D(T= 0) = \l\{\begin{array}{ll}
 n,& \eta \to 0 \\
\frac{n+1}{\pi} \,\sin\frac{n \pi}{n+1},& \eta \to \infty 
\end{array}\r. \label{special}\,.
\ee
For the latter case, the results for the parameters $v_c,\,K_c$, as calculated in \cite{borser06}, have been inserted.  

Secondly, the results in \refeq{chic} are not specific for the $q$-deformed Bose model considered in this paper, but are valid for all models that, in the low-energy region, are described by the Gaussian model \refeq{gauss2}. Let us give two examples. 

The first example is the Lieb-Liniger Bose gas \refeq{nls}. As shown by Haldane \cite{hal81}, its low-energy properties are equally described by the Gaussian model \refeq{gauss2}. Especially, the relations \refeq{vck}, valid in the small-coupling regime, are the same for the Bose gas \refeq{nls} for weak coupling $c$. Thus both models are equivalent in the weak-coupling limit. In the strong-coupling regime of the Bose gas, however, $v_c$ and $K_c$ have to be calculated from the Bethe ansatz solution \cite{ll63,korbook}. 

As a second example, consider the critical spin-1/2 $XXZ$-Heisenberg chain. In
this case, the low-energy theory is obtained by performing a Jordan-Wigner
transformation of the spin operators, resulting in a lattice model of spinless
fermions. In the continuum limit, the corresponding fermionic operators are written in terms of bosonic fields (bosonization), resulting in the Hamiltonian density \cite{luk98,fuj03}
\be
H_0&=& \frac{v_s}{2} \l[ \Pi_s^2 + \l(\partial_x \phi_s\r)^2\r]\nn\\
H_{XXZ}&=&H_0 - h \sqrt{\frac{K_{s}}{\pi}} \partial_x \phi_s - \tilde \lambda \sqrt{\frac{K_{s}}{\pi}} v_s \Pi_s\label{xxz},
\ee
with $\l[\phi_s(x),\Pi_s(y)\r] = \rmi \delta(x-y)$. Here, $v_s$ is the velocity of spin excitations and $K_s$ the corresponding Luttinger liquid parameter. In \refeq{xxz}, two terms additional to $H_0$ have been included: 
The magnetic field $h$ couples to the spin density (stemming from the spin operator $S_\ell^z$ in the lattice model), which yields the fermionic density. Secondly, $\tilde \lambda$ couples to the spin, i.e. fermionic current, originating in the operator $(S_{\ell}^+S_{\ell+1}^- - S_{\ell}^- S_{\ell+1}^+)/(2\rmi)$ in the lattice model. By comparing with Eq.~\refeq{gauss2}, one has $\Phi'=\partial_x\phi'\propto \partial_x \phi_s$ and $\partial_x \theta'\propto \Pi_s$. We now set $\phi_s=\phi_{L} + \phi_{R}$, $\Pi_s=:\partial_x \theta_s=\partial_x\l(\phi_L - \phi_R\r)$, resulting in essentially the same Hamiltonian density \refeq{lrmod} (apart from trivial prefactors in the potentials $\mu, \tilde \lambda$). Thus the magnetic susceptibility $\chi_s$ and the Drude weight $D_s$ of the spin current are given by $\chi_s(T=0)\propto K_s/v_s$ \cite{luk98}, $D_s(T=0)\propto v_s K_s$ \cite{sha90,fuj03}, with trivial constants of proportionality. 

We finally remark on next-leading contributions to the leading low-temperature result \refeq{dsmallt}. Such contributions are determined by irrelevant operators in the effective field theory, which have been neglected in \refeq{h0}. In the case of the $q$-Bose model, these stem from 
\begin{itemize}
\item[i)] higher order-terms in the series expansions (\ref{xopapp},\ref{yopapp}),
\item[ii)] corrections to the long-wavelength approximations (\ref{xopapp},\ref{yopapp}) due to the discrete nature of the particle density.
\end{itemize}
As pointed out in \cite{hal81,gia92}, the second kind of corrections can, in
lowest order, principally give rise to a term $\propto \cos 2\pi(\phi + n
x/\Delta)$. Due to the lattice, such a term is commensurate at density
$n=1$. After the scaling \refeq{scal}, we see that this term becomes relevant
for $K_c< 2/\pi$. The analysis in \cite{borser06} shows that $K_c\in
\l]1/\pi,\infty\r[$, so that a massive phase seems to be possible. Such a
    scenario has indeed been identified in the generic Bose-Hubbard model
    \cite{kue00}. In our case, however, this would contradict the conservation
    of the current, Eq.~\refeq{com}: Since $\l[\exp\l[\rmi
	\phi(x)\r],J(x)\r]=\delta(x-y) n \exp\l[\rmi \phi(x)\r]$, the above
    $\cos$-term cannot occur at all.\footnote{The same is true for higher order terms of the form $\cos 2 \pi m \phi$.}  This leads us to the conclusion that terms due to i) in the above list cancel those due to ii). It would be desirable to show this explicitly, a goal which, at this point, is left for future work.  

Further candidates for irrelevant operators have been given  in \cite{borser06}. These have even scaling dimensions, so that the Drude weight behaves as
\be
D(T\gtrsim 0)=v_c K_c  + \mathcal O \l( T^2\r)\label{nextdct}\,.
\ee
The amplitude of the next-leading term is unknown so far and given by the amplitudes of the leading irrelevant operators. 

It is interesting to note that the $\cos$-terms due to ii) can be ``fermionized'' and are then interpreted as Umklapp-scattering processes. Such processes are present in the $XXZ$-chain and accordingly the corresponding operators occur as irrelevant contributions in the low-energy effective theory. Additionally, operators due to the band curvature at the Fermi points are present. Both types of operators lead to temperature-dependent corrections to $D_s(T=0)$, with non-integer and integer powers, respectively \cite{fuj03,js06}. Thus although both models are, in the leading order, equivalently described by \refeq{lrmod}, they differ in the irrelevant operators. 
\section{Thermodynamic Bethe Ansatz}
In this section, an exact formula for the temperature dependence of $D(T)$ is presented. This formula is evaluated analytically in the low- and high-temperature limits, as well as numerically for a wider range of temperatures. 

\subsection{Low- and high-temperature limits}
Employing the TBA originally developed by Yang and Yang to calculate the thermodynamics of the Lieb-Liniger Bose gas \cite{yy69}, the grand-canonical potential of the model \refeq{defh} has been obtained \cite{bul95} and analyzed \cite{borser06}. Analogously, the thermodynamic potential $g(T,\mu)$ corresponding to \refeq{wth} with $\tilde \la=\la/\beta$ reads
\be
-\beta g(T,\mu)&=& \int_{-\pi}^\pi \ln A(k) \frac{\d k}{2\pi}\label{betag},
\ee
with $a:=1+A$ and
\be
\ln a(k)& =& \beta \cos k + \beta \mu + \la \sin k +  \int_{-\pi}^\pi \kappa(k-q) \ln A(q) \frac{\d q}{2\pi}\label{loga}\\
\kappa(k)&=&\frac{\sinh 2\eta}{\cosh 2\eta - \cos k}\nn.
\ee
Then, following  Eqs.~(\ref{dt},\ref{jsqu}),
\be
D(T)&=& \beta\,\partial^2_\la \l.\beta g(T,\mu)\r|_{\la=0,n=const.} \label{tbadt}\,.
\ee
Note that the derivatives with respect to $\la$ are to be taken such that the particle density $n$ is held constant. 

Before diving into the analysis of the low- and high-temperature limits, it is useful to envision the following relations:
\be
\partial_\la \ln A(k)& = &\frac{\partial_\la \ln a(k)}{1+\exp[-\ln a(k)]}\label{aux1}\\
\partial^2_\la \ln A(k)&=& \frac{\partial^2_\la \ln a(k)}{1+\exp[-\ln a(k)]}+ \frac{\l[\partial_\la \ln a(k)\r]^2}{4 \cosh^2 \frac{\ln a(k)}{2}}\;.\label{aux2}
\ee
Let us now first evaluate Eq.~\refeq{tbadt} in the low-temperature limit. In this case, for $\la=0$,  
\be
\ln a(k)&=& -\beta \ve_0(k)+\mathcal O\l(T\r)\label{lna}\\
\ve_0(k)& =& -\cos k-\mu + \int_{-B}^B \kappa(k-q) \ve_0(q) \frac{\d q}{2\pi}\nn\\
\ve_0(\pm B) &=& 0,\; \ve_0(|k|<B)<0 \label{ve}
\ee
We neglect the next-leading correction $\mathcal O(T)$ in Eq.~\refeq{lna} in the ongoing. Then
\be
\partial_\la \ln a(k) &=& \sin k + \int_{-\pi}^\pi \kappa(k-q) \partial_\la \ln A(q) \frac{\d q}{2\pi} \nn\\
&\stackrel{T\to 0}{\to}& \sin k + \int_{-B}^B \kappa(k-q) \partial_\la \ln a(q) \frac{\d q}{2\pi}\nn,
\ee
where in the last line, we used Eqs.~(\ref{aux1},\ref{lna},\ref{ve}). Obviously, 
\be
\lim_{T\to 0} \partial_\la \ln a(k) = \partial_k \ve_0(k) \label{lalna}.
\ee 
Next, we consider the equation for $\partial^2_\la \ln a(k)$, by making use of Eq.~\refeq{aux2}:
\be
\fl \partial^2_\la \ln a(k) = \int_{-\pi}^\pi \kappa(k-q) \partial_\la^2 \ln A(q) \frac{\d q}{2\pi}\nn\\
\fl \qquad \stackrel{T\to 0}{\to} T \partial_k \ve_0(B)\l[ \kappa(k-B) + \kappa(k+B)\r] + \int_{-B}^B \kappa(k-q) \partial^2_\la \ln a(q) \frac{\d q}{2\pi}\label{lnall}.
\ee
The last term in the last line is obtained from the first term in \refeq{aux2}, using Eqs.~\refeq{lna},\refeq{ve}. To obtain the second term, we performed the approximation
\be
\fl\lim_{\beta\to \infty}\frac{1}{4\cosh^2 \frac{\ln a(k)}{2}} =\lim_{\beta\to \infty}\frac{1}{4\cosh^2 \frac{\beta \ve_0(k)}{2}}= \frac{1}{\partial_k \ve_0(B)}\l[ \delta(k-B) + \delta(k+B)\r]\nn
\ee
where use was made of Eq.~\refeq{lalna}. One now observes that 
\be
\int_{-B}^B \partial_\la^2 \ln a(q) \frac{\d q}{2\pi} &=& T \partial_k \ve_0(B)\l[ -\frac{1}{\pi} + 2 \rho(B)\r]\label{eqrho},
\ee
where the function $\rho(k)$ is given by
\be
\rho(k) &=& \frac{1}{2\pi} + \int_{-B}^B \kappa(k-q) \rho(q) \frac{\d q}{2 \pi}\nn.
\ee
In fact, $\rho(|k|\leq B)$ describes the density of BA-roots in the interval $\l[-B,B\r]$. Eq.~\refeq{eqrho} can be established by writing down the integral series for the left and right hand sides of the equation and comparing term by term. Combining \refeq{tbadt} with \refeq{betag} and \refeq{lnall}, we thus arrive at the result
\be
D(T\gtrsim 0)=2  \rho(B) \partial_k \ve_0(B)  \label{dsmallt2}
\ee
Now we recall how the charge velocity $v_c$ and the Luttinger parameter $K_c$ are obtained within the BA-solution \cite{korbook,borser06}: 
\be
v_c&=& \frac{\partial_k \ve_0(B)}{2\pi \rho(B)}\nn\\
K_c&=&4\pi \rho^2(B)\nn,
\ee
which confirms that Eq.~\refeq{dsmallt} is valid for the entire range $\eta\in\l]0,\infty\r[$. 

We now consider the high-temperature limit $T\to \infty$. In this case, Eq.~\refeq{loga} simplifies to
\be
\ln a(k)= \beta \mu + \int_{-\pi}^\pi \kappa(k-q) \ln A(q) \frac{\d q}{2\pi} \label{lnaht},
\ee
which is solved by 
\be
\ln a=-\ln \l(\te^{-\beta \mu} -1\r)\label{lnaconst},
\ee
without any dependence on the spectral parameter $k$ (note that $\int_{-\pi}^\pi \kappa(k-q) \d q = 2\pi$). In order to derive Eq.~\refeq{lnaht}, it was essential that the $k$-dependent driving term in Eq.~\refeq{loga} is bounded, a fact due to the presence of the lattice. In the analogous equation for the continuous Bose-gas, the driving-term is unbounded \cite{yy69}, which up to now makes it impossible to write down a high-temperature expansion for that model. 

Taking into account Eqs.~(\ref{aux1},\ref{lnaconst}), we obtain the equation for $\partial_\la \ln a(k)$, 
\be
\partial_\la \ln a(k) = \sin k + \te^{\beta \mu} \int_{-\pi}^\pi  \kappa(k-q) \partial_\la \ln a(q) \frac{\d q}{2\pi} \label{lalnaht}.
\ee
It is solved explicitly for $\eta \to \infty$ and $\eta \to 0$: In the $\eta\to\infty$-case, $\kappa\equiv 1$ and $\partial_\la \ln a(k) = \sin k$. In the other extreme, $\eta\to 1$, $\kappa(k)=2\pi \delta(k)$ and thus $\partial_\la \ln a(k)=\sin k/(1-\exp\l[\beta \mu\r])$. 

From Eq.~\refeq{aux2} we calculate $\int_{-\pi}^\pi \partial^2_\la \ln a(k) \frac{{\mbox{{\footnotesize d}}} k}{2\pi}$, again using Eq.~\refeq{lnaconst}. We find
\be
\int_{-\pi}^\pi \partial^2_\la \ln a(k) \frac{\d k}{2\pi}= \frac{1+a}{4\cosh^2 \frac{\ln a}{2}} \, \int_{-\pi}^\pi \l[ \partial_\la \ln a(q)\r]^2 \frac{\d q}{2\pi} \nn
\ee
and therefore 
\be
\lim_{T\to \infty} D(T)&=&\beta \lim_{T\to \infty}\int_{-\pi}^\pi \partial^2_\la \ln A(q) \frac{\d q}{2\pi} \nn\\
&=& \frac{(a+1)\beta}{4\cosh^2 \frac{\ln a}{2}} \,\int_{-\pi}^\pi \l[ \partial_\la \ln a(q)\r]^2 \,\frac{\d q}{2\pi}\label{dht}\,.
\ee

Eq.~\refeq{dht} gives the high-temperature Drude weight as a function of
$\beta \mu$, via Eqs.~(\ref{lalnaht},\ref{lnaconst}). Actually, we are interested in $D(T)$ at constant particle density $n$, 
\be
n=-\partial_\mu g = T \int_{-\pi}^\pi \frac{\partial_\mu \ln a(q)}{1+\exp\l[-\ln a(q)\r]} \frac{\d q}{2\pi}.\nn
\ee
From the integral equation for $\partial_\mu \ln a(k)$ it follows, using Eq.~\refeq{lnaconst}, that $\int_{-\pi}^\pi \partial_\mu \ln a(q) \frac{\mbox{{\footnotesize d}} q}{2\pi} = \beta(1+a)$, so that for $T\to \infty$, 
\be
n=a&=&\frac{1}{\te^{-\beta \mu} -1}\nn\\
\mu(T\to \infty) &=& -T \ln \frac{1+n}{n}\nn.
\ee
With these findings, we rewrite Eq.~\refeq{dht} as
\be
\lim_{T\to \infty} D(T)&=&\frac{(1+n)\beta}{2+n+n^{-1}} \,\int_{-\pi}^\pi \l[ \partial_\la \ln a(q)\r]^2 \,\frac{\d q}{2\pi}\label{dht2}.
\ee
Especially, for $\eta\to \infty$ and $\eta \to 0$,
\be
\lim_{T\to \infty} D(T)&=&\l\{\begin{array}{ll} 
\frac{1+n}{2(2+n+n^{-1})}\,\beta,& \eta \to \infty \\
\frac{(1+n)^3}{2(2+n+n^{-1})}\,\beta,& \eta \to 0 
\end{array}\r. \label{dht3}.
\ee

\subsection{Numerical results}
\begin{figure}
\begin{center}
\includegraphics*[scale=0.5]{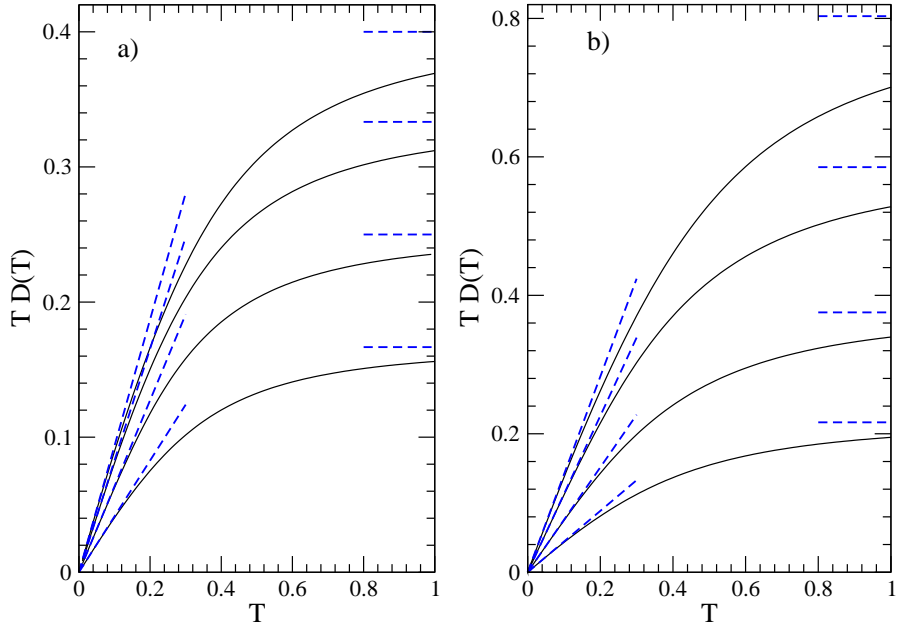}
\caption{The Drude weight evaluated numerically from Eq.~\refeq{tbadt}, for a) $q=0$ and b) $q=\exp[-0.5]$. The curves show the results for $n=0.5, 1, 2, 4$ (from bottom to top). The dashed lines near $T=0$ and $T=1$ are the low- and high-temperature limits, Eqs.~(\ref{special},\ref{dsmallt2}) and (\ref{dht},\ref{dht3}), respectively.} 
\label{fig1}
\end{center}
\end{figure}

\begin{figure}
\begin{center}
\includegraphics*[scale=0.5]{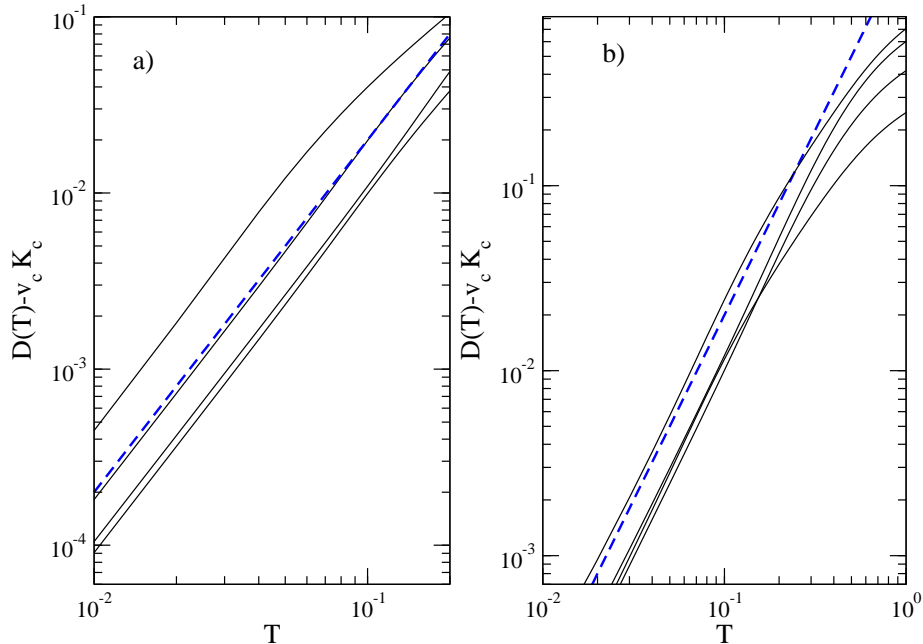}
\caption{The quantity $D(T)-v_c K_c$ at low temperatures on a double logarithmic scale, for a) $q=0$ and b) $q=\exp[-0.5]$. The curves show the results for $n=0.5, 1, 2, 4$ (from bottom to top at the highest temperatures shown). For comparison, the dashed lines show the function $2\cdot T^2$. In all cases, the exponent of the plotted quantity equals 2, independent of the density and the interaction parameter, thereby confirming Eq.~\refeq{nextdct}.} 
\label{fig2}
\end{center}
\end{figure}

The Drude weight $D(T)$ is calculated numerically according to Eq.~\refeq{tbadt}, by setting up separate integral equations for the partial derivatives. The convolution integrals are treated using the Fast Fourier Transform algorithm. The chemical potential is adjusted for all temperatures considered in order to keep $n$ at a fixed value. Results for $q=0$ and $q=\exp[-0.5]$ are shown in Fig.~\ref{fig1}. Furthermore, in Fig.~\ref{fig2} the next-leading contribution $D(T)-v_c\,K_c$ at low temperatures is plotted, yielding evidence that $ D(T)- v_c\,K_c=\mathcal O\l(T^2\r)$, in agreement with Eq.~\refeq{nextdct}.

\section{Conclusion and outlook}
We have calculated exactly the finite-temperature Drude weight $D(T)$ of a one-dimensional bosonic lattice system. Due to the conserved particle current, $D(T)$ is finite. Note that the converse is not true: In the critical spin-1/2 $XXZ$ chain, the current is not conserved, but $D(T)$ is finite nevertheless. 

At low temperatures, agreement was found between the results derived from an effective Gaussian field theory and the BA. At high-temperatures, $D(T)$ approaches a finite value which was calculated analytically from the BA solution as well. A numerical solution of the TBA-equations extrapolates between the low- and high-temperature limits. 

Open questions for future work are threefold: First, it would be interesting to show the absence of ``Umklapp''-like terms in the $q$-Bose model explicitly. Next, the amplitudes of the next-leading terms of $D(T)$ in the low-temperature regime are of intrinsic theoretical interest. They are directly linked to the amplitudes of irrelevant operators in the effective field theory. Finally, after the particle current, the thermal current in the Bose chain is an interesting object for future studies. It is known that in other integrable models, like, for example, in the spin-1/2 $XXZ$-chain, the thermal current is conserved \cite{zot97}, resulting in a finite thermal Drude weight at finite temperatures. In how far this also holds for the $q$-Bose chain is to be explored in future work. 

\section*{Acknowledgments}
The author thanks M. Gulasci, S. Sergeev and J. Sirker for helpful discussions. This work has been supported by the German Research Council (DFG) under grant number BO 2538/1-1. 


\end{document}